\documentclass{llncs}
\usepackage{amssymb}
\usepackage{pxfonts}
\newcommand{\sll}{\mathit{SLL}}
\newcommand{\dagg}{\mathit{DAG}}
\newcommand{\tr}{\mathit{T}}
\newcommand{\cy}{\mathit{C}}
\newcommand{\var}{\mathit{Var}}
\newcommand{\lay}{\mathit{layout}}
\begin{document}
\title{Recognition of Logically Related Regions Based Heap Abstraction}
\author{Mohamed A. El-Zawawy} \institute{College of Computer and
Information Sciences\\
Al-Imam M. I.-S. I. University\\ Riyadh 11432\\Kingdom of Saudi
Arabia\\ and \\Department of
Mathematics\\ Faculty of Science \\ Cairo University\\ Giza 12613\\
Egypt\\ \email{maelzawawy@cu.edu.eg}} \maketitle
%----------------------------------------------------------------------------------------
%----------------------------------------------------------------------------------------
\begin{abstract}
This paper presents a novel set of algorithms for heap abstraction,
identifying logically related regions of the heap. The targeted
regions include objects that are part of the same component
structure (recursive data structure). The result of the technique
outlined in this paper has the form of a compact normal form (an
abstract model) that boosts the efficiency of the static analysis
via speeding its convergence. The result of heap abstraction,
together with some properties of data structures, can be used to
enable program optimizations like static deallocation, pool
allocation, region-based garbage collection, and object co-location.

More precisely, this paper proposes algorithms for abstracting heap
components with the layout of a singly linked list, a binary tree, a
cycle, and a directed acyclic graph. The termination and correctness
of these algorithms are studied in the paper. Towards presenting the
algorithms the paper also presents concrete and abstract models for
heap representations.
\end{abstract}
\footnote{Short title: Recognition of Logically Related Regions
Based Heap Abstraction. }
%----------------------------------------------------------------------------------------
%----------------------------------------------------------------------------------------
\section{Introduction}\label{intro}
This paper presents an efficient technique for heap abstraction
which takes the form of identifying and grouping logically related
regions of heaps. The result of heap abstraction is a normal form
for the program heap. The normal form is necessary for abstractly
modeling programs and it boosts the efficiency of the static data
flow analysis via assisting the analysis to converge faster. The
information provided by the normal form can also be used by client
optimization applications to achieve the analyses of object
co-location, pool allocation~\cite{Wang10}, static
deallocation~\cite{Dios10}, etc.

The concept of heap abstraction emerges naturally in the course of
research on object allocation and memory layout where techniques
like pool allocation and object co-location use the heap abstraction
to improve object locality. The efficiency of garbage
collection~\cite{Schoeberl10} is also boosted using heap abstraction
by the means of other techniques. Applications that statically
deallocate data-structures or regions use heap abstractions more
directly than others. In the course of abstraction, by restricting
the grouping process to regions of heap that are expected to contain
dead objects, the abstracting information can be used to delay times
of garbage collection. Parallel garbage collection can also be
treated by other approaches that use heap write/read information to
statically find regions of heap that can be securely grouped without
burden for the mutator.

Various techniques for heap
abstraction~\cite{Cherem06,Magill10,Marron09} are used by the
approaches referred to above to get region information used later in
the optimization stage. The simplest of these approaches groups the
heap objects based on the result of a pointer analysis like
Steensgaard analysis~\cite{Steensgaard96}. Most of these techniques
do not conveniently model object-oriented properties of data
structures because the techniques are based on pointer analysis or
other analyses that are flow/context insensitive. This paper
presents a technique which is much more precise than these
techniques. Moreover, our technique can be used as a tool to
optimize the heap, which results in boosting the efficiency of
memory regions.

Additionally, our proposed technique is useful to improve the
efficiency of a range of static analysis approaches. This is
accomplished via using the heap abstraction to normalize abstract
models~\cite{Beek11} which in turn results in reducing the height of
the abstract lattice. Therefore this normalization process can be
realized as a widening operation that turn domains of infinite
height into ones of finite height. The normalization mentioned here
has two aspects. One aspect is the compactification of recursive
structures of possibly infinite size into finite structures. The
other aspect is the using of a similarity relation to group objects,
making up composite data structures.

Although the idea of heap abstraction (normalization) has already
appeared in existing research, the algorithms presented in this
paper achieving heap abstraction are more general, precise, and
reliable (their correctness are proved) than those that have been
developed in previous works. Precisely, our approach applies to a
variety of recursive data structures such as singly linked lists,
binary trees, cycles, and acyclic directed graphs. The technique
presented in this paper can appropriately handle multi-component
structures which could not be handled by most existing works.

\subsection*{Motivating example}
Figures~\ref{sll},~\ref{bt},~\ref{cycle}, and~\ref{DAG} present
motivating examples for our work. Suppose that we have a heap whose
cells have the shape of four components; the singly linked list, the
binary tree, the cycle, and the DAG (abbreviation for directed
acyclic graph) on the left-hand side of
Figures~\ref{sll},~\ref{bt},~\ref{cycle}, and~\ref{DAG},
respectively. We note that some cells like \begin{enumerate}
    \item the nodes pointed to by variables $s$ and $e$ in the linked list, the
cycle, and the DAG.
    \item the second last node of the linked list, and
    \item the root of the binary tree and the third and
fourth (from left to right) nodes of the third level of the binary
tree
\end{enumerate} are interesting and contain additional information
as compared to the remaining cells. Other cells of the heap, like
the second and third nodes of the linked list, are ordinary and
carries no extra information. It is wise and helpful to abstract
such heap into one that consists of the four components on the right
hand side of Figures~\ref{sll},~\ref{bt},~\ref{cycle},
and~\ref{DAG}. The meant abstraction here is that of grouping
logically related cells of the heap.
\begin{remark}
Self edges in abstracted components of the heap of our example have
different meanings depending on the component layout.
\end{remark}
\subsection*{Contributions}
Contributions of this paper are the following:
\begin{enumerate}
\item A new technique for heap abstraction; novel algorithms for
identifying and grouping logically related cells in singly linked
lists, binary trees, cycles, and directed acyclic graphs. The
termination and correctness of these algorithms are studied.
\item New concrete and abstract models for heap representations;
a formal concept (valid abstraction) capturing the relationship
between a concrete model and its abstraction.
\end{enumerate}

\subsection*{Organization}
The rest of the paper is organized as follows. Section~\ref{model}
presents the parametrically labeled storage shape graph models
(concrete and abstract) that we use to describe our new technique
for heap abstraction.
Sections~\ref{sll-s},~\ref{bt-s},~\ref{cycle-s}, and~\ref{dag-s}
present new algorithms for identifying and grouping logically
related cells in singly linked lists, binary trees, cycles, and
directed acyclic graphs, respectively. The algorithm that abstracts
heaps and that calls other introduced algorithms is presented in
Section~\ref{heap-s}. Related work is briefly reviewed in
Section~\ref{related work}.

\section{Concrete and abstract heap models}\label{model}
This section introduces heap models that the work presented herein
builds on. Graphs are basic components of our models. Similar models
are used in related work~\cite{Marron09,Marron08} towards shape and
sharing analysis of Java programs. It is worth mentioning that
concepts of this paper are applicable in other techniques based on
separation logic~\cite{Berdine07,Jeannet10,Yang08}.

As usual, our semantics of memory  is defined using pairs of stacks
and heaps. The stack assigns values to variables and the heap
assigns values to memory addresses. Each pair of a stack and a heap
is called a \textit{concrete component}. The concept of
\textit{concrete heap} denotes a finite set of concrete components.
A concrete component is represented by a labeled directed graph
which has a layout attribute that captures the layout of the memory
cells represented by the component. The precise definition is the
following:

\begin{definition} A \textit{concrete heap} is
a finite set of disjoint labeled directed graphs (called
\textit{concrete components}) $C_1,\ldots,C_n$ each of which has a
\textit{layout} attribute that can have one of four values; singly
linked list ($\sll$), tree ($\tr$), cycle ($\cy$), and directed
acyclic graph ($\dagg$). The layout of a component, $C_i$, is
denoted by $C_i.\lay$. More precisely, $C_i=(V_i,A_i,P_i)$, where:
\begin{enumerate}
\item $V_i$ is a finite set of variables; $V_i\subseteq \var$.
\item $A_i$ is a finite set of memory addresses; $A_i\subseteq \textit{Addrs}$.
\item When $C_i.\lay = \sll, \dagg, \mbox{or }
\cy$, $P_i$ is a set of pointers defined by ${P_i \subseteq
(V_i\times A_i)\cup (A_i\times A_i)}$.
\item When $C_i.\lay = \textit{T},\ P_i
\subseteq (V_i\times A_i)\cup (A_i\times A_i\times \{l,r\})$.
\end{enumerate}
\end{definition}

Regions in heaps, edges of regions, edges entering regions, and
edges leaving regions are defined as follows:
\begin{definition}
A set $R$ is said to be a \textit{region} in a concrete component
$C=(V,A,P)$ if $R\subseteq A$. Moreover,
\begin{enumerate}
\item $P(R)=\{(a_1,a_2,nx),(a_1,a_2)\in P\mid a_1,a_2\in R\},$
\item $P_\textit{in}(R)=\{(a_1,a_2,nx),(a_1,a_2)\in P\mid
a_1\in A\setminus R,a_2\in R\},$ and
\item $P_\textit{out}(R)=\{(a_1,a_2,nx),(a_1,a_2)\in
P\mid a_1\in R,a_2\in A\setminus R\}$.
\end{enumerate}
\end{definition}

Our concept of abstract heap is inspired by the technique of storage
shape graph presented in~\cite{Chase90,Marron09}. The concept of
concrete component is abstracted by that of abstract component which
is a labeled directed graph $(\hat{V},\hat{N},\hat{P})$, where (a)
$\hat{V}$ is a set of nodes correspond to variables, (b) $\hat{N}$
is a set of nodes each of which corresponds to (abstracts) a region
of a concrete component, and (c) $\hat{P}$ is a set of graph edges,
each of which corresponds to (abstracts) a set of pointers.
Analogously to concrete component, each abstract component has a
layout attribute. More precisely abstract heaps and abstract
components are defined as follows:

\begin{definition}
An \textit{abstract heap} is a finite set of disjoint labeled
directed graphs (called \textit{abstract components})
$\hat{C}_1,\ldots,\hat{C}_n$ each of which has a \textit{layout}
attribute that is $\sll, \tr, \cy$, or $\dagg$. More precisely,
$\hat{C}_i=(\hat{V}_i,\hat{N}_i,\hat{P}_i)$ where:
\begin{enumerate}
\item $\hat{V}_i\subseteq \var$.
\item $\hat{N}_i$ is a finite set of node identifiers (each represents a region of the heap).
\item When $\hat{C}_i.\lay = \sll, \dagg, \mbox{or }
\cy$, $P_i$ is a set of pointers defined by ${\hat{P}_i \subseteq
(\hat{V}_i\times N_i)\cup (N_i\times N_i)}$.
\item When $\hat{C}_i.\lay = \textit{T},\ P_i
\subseteq (\hat{V}_i\times N_i)\cup (N_i\times N_i\times \{l,r\})$.
\end{enumerate}
\end{definition}

Regions in abstract components are defined analogously as sets of
nodes identifiers.

\begin{remark}
Every concrete heap is an abstract one.
\end{remark}

Now we introduce the concept of abstraction. An abstract component
$\hat{C}$ is described as a valid abstraction of another one
$\hat{C}^\prime$, if (a) they have the same layout and same sets of
variables and (b) there are two maps; a map from nodes of $\hat{C}$
to nodes of $\hat{C}^\prime$ and a map from edges of $\hat{C}$ to
edges of $\hat{C}^\prime$ such that these maps preserve the
connectivity of the components.

\begin{definition}
An abstract component $\hat{C}_1=(\hat{V}_1,\hat{N}_1,\hat{P}_1)$ is
a \textit{valid abstraction} of another  abstract component
$\hat{C}_2=(\hat{V}_2,\hat{N}_2,\hat{P}_2)$ if
$\hat{C}_1.\lay=\hat{C}_2.\lay,\ \hat{V}_1=\hat{V}_2$, and there are
two onto maps $f_N:\hat{N}_1\longrightarrow \hat{N}_2\mbox{ and }
f_P:\hat{P}_1\longrightarrow \hat{P}_2$ such that:
\begin{enumerate}
\item $\forall (v,n_2)\in \hat{P}_2.\ f_P^{-1}((v,n_2))
\subseteq \{(v,n_1)\in \hat{P}_1\mid n_1\in f_N^{-1}(n_2)\}$.
\item $\forall (n_2,n_2^\prime)\in \hat{P}_2.\ f_P^{-1}((n_2,n_2^\prime))
\subseteq \{(n_1,n_1^\prime)\in \hat{P}_1\mid n_1\in
f_N^{-1}(n_2)\wedge n_1^\prime\in f_N^{-1}(n_2^\prime)\}$.
\item $\forall (n_2,n_2^\prime,nx)\in \hat{P}_2.\ f_P^{-1}((n_2,n_2^\prime,nx))
\subseteq {\{(n_1,n_1^\prime,nx)\in \hat{P}_1\mid n_1\in
f_N^{-1}(n_2)\wedge n_1^\prime\in f_N^{-1}(n_2^\prime)\}}$.
\end{enumerate}
The pair $(f_N,f_P)$ is called the witness of the valid abstraction.
\end{definition}

\begin{lemma}\label{lem1}
The valid-abstraction relation on abstract components is transitive.
\end{lemma}
\begin{proof}
Suppose $\hat{C}_2$ is a valid abstraction of $\hat{C}_1$ with
witness $(f_N,f_P)$ and $\hat{C}_3$ is a valid abstraction of
$\hat{C}_2$ with witness $(f_N^\prime,f_P^\prime)$. Then, it is easy
to see that $ {(f_N^\prime\circ f_N,f_P^\prime\circ f_P)}$ witnesses
that $\hat{C}_3$ is a valid abstraction of $\hat{C}_1$.
\end{proof}

\begin{definition}
An abstract heap $(\hat{C}_1,\ldots,\hat{C}_n)$ is a valid
abstraction of a concrete heap $(C_1,\ldots,C_n)$ if for every $1\le
i\le n,\ \hat{C}_i $ is a valid abstraction of $C_i$.
\end{definition}

\begin{remark}
Every concrete heap is a valid abstraction of itself.
\end{remark}

\begin{figure}[htb]
\unitlength.75mm
\begin{center}
\begin{picture}(112,10)(0,0)
\put(-10,10){\oval(7,5)} \put(-6.5,10){\vector(1,0){5}}
\put(-9,11){\makebox[0pt]{\raisebox{-1ex}{{\small $h_0$ }}}}
\put(-10,0){\circle{5}}
\put(-9,1){\makebox[0pt]{\raisebox{-1ex}{{\small s }}}}
\put(-10,2.5){\vector(0,1){5}}
\put(2,10){\oval(7,5)}\put(5.5,10){\vector(1,0){5}}
\put(3,11){\makebox[0pt]{\raisebox{-1ex}{{\small $h_1$ }}}}
\put(14,10){\oval(7,5)} \put(17.5,10){\vector(1,0){5}}
\put(15,11){\makebox[0pt]{\raisebox{-1ex}{{\small $h_2$ }}}}
\put(26,10){\oval(7,5)}\put(29.5,10){\vector(1,0){5}}
\put(27,11){\makebox[0pt]{\raisebox{-1ex}{{\small $h_3$ }}}}
\put(38,10){\oval(7,5)} \put(41.5,10){\vector(1,0){5}}
\put(39,11){\makebox[0pt]{\raisebox{-1ex}{{\small $h_4$ }}}}
\put(50,10){\oval(7,5)}\put(53.5,10){\vector(1,0){5}}
\put(51,11){\makebox[0pt]{\raisebox{-1ex}{{\small $h_5$ }}}}
\put(62,10){\oval(7,5)}
\put(63,11){\makebox[0pt]{\raisebox{-1ex}{{\small $h_6$ }}}}
\put(65.5,11){\vector(1,0){5}}\put(70.5,9){\vector(-1,0){5}}
\put(74,10){\oval(7,5)} \put(74,0){\circle{5}}
\put(75,11){\makebox[0pt]{\raisebox{-1ex}{{\small $h_7$ }}}}
\put(75,1){\makebox[0pt]{\raisebox{-1ex}{{\small e }}}}
\put(74,2.5){\vector(0,1){5}}
%----------------------------------------------------------------
\put(90,10){\oval(7,5)} \put(93.5,10){\vector(1,0){5}}
\put(91,11){\makebox[0pt]{\raisebox{-1ex}{{\small $i_0$ }}}}
\put(90,0){\circle{5}}
\put(91,1){\makebox[0pt]{\raisebox{-1ex}{{\small s }}}}
\put(90,2.5){\vector(0,1){5}} \put(102,10){\oval(7,5)}
\put(103,11){\makebox[0pt]{\raisebox{-1ex}{{\small $i_1$ }}}}
\put(100,12.5){{\large $\curvearrowright$}}
\put(105.5,10){\vector(1,0){5}} \put(114,10){\oval(7,5)}
\put(115,11){\makebox[0pt]{\raisebox{-1ex}{{\small $i_2$ }}}}
\put(117.5,11){\vector(1,0){5}}\put(122.5,9){\vector(-1,0){5}}
\put(126,10){\oval(7,5)} \put(126,0){\circle{5}}
\put(127,11){\makebox[0pt]{\raisebox{-1ex}{{\small $h_3$ }}}}
\put(126,1){\makebox[0pt]{\raisebox{-1ex}{{\small e }}}}
\put(126,2.5){\vector(0,1){5}}
\end{picture}
\end{center}
\caption{A concrete singly linked list together with its
abstraction.} \label{sll}
\end{figure}
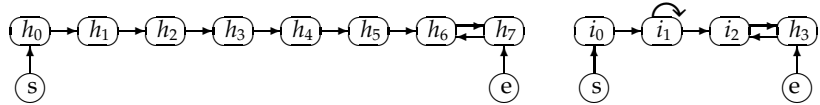
\section{Abstracting singly linked lists}\label{sll-s}
This section presents a novel algorithm for abstracting heap
components whose layouts are singly linked list. Figure~\ref{sll}
presents a linked list of length 8 (left) together with its
abstracted representation (right) which is a list of length 4.
Clearly, some nodes of the concrete list are grouped into regions.
We note that nodes $h_0$ and $h_7$ are special nodes as they are
pointed to by two variables s and e, respectively. The node $h_6$ is
also special as it shares a back edge with the node $h_7$. These
special nodes are not grouped in the abstracted version of the list.
The nodes $h_1$ up to the node $h_5$ are ordinary nodes; there is
nothing special about them. These ordinary nodes are grouped into
the node (region) $i_1$ in the abstracted version. The self-edge of
$i_1$ captures the phenomenon that $i_1$ represents a region of the
concrete component.

Various properties of lists are captured by partitioning the list
nodes into two classes; ordinary and special nodes. First, the
compressed representation of the list in the abstracted version
substantially boosts the efficiency of the analysis. Next, the first
and last nodes, pointed to by variables $s$ and $e$ respectively,
and the second-to-last node are kept separate. This supports the
analysis to conveniently simulate the semantics of later program
commands. Although ordinary nodes of each list in the program are
grouped into a single node in the abstracted version of the list,
unrelated lists of the program are kept separate. In other words,
separate lists in the concrete heap are kept separate in the
abstract model while nodes in the same lists are grouped together.
This helps in preserving the information required by many
optimization techniques. The formal definition of special and
ordinary nodes of singly linked lists is as follows:

\begin{definition}
Let $\hat{C}=(\hat{V},\hat{N},\hat{P})$ be an abstract component
whose layout is $\sll$. Then, a node $n\in \hat{N}$ is
\textit{special} if either:
\begin{enumerate}
    \item for some $\hat{v}\in \hat{V},\ (\hat{v},n)\in \hat{P}$, or
    \item there exists $(a,b)\in \hat{P}$ such that
    $a,b\in \hat{N},\ \textit{depth}(a)> \textit{depth}(b)$, and $n\in\{a,b\}$;
    i.e.,
    $n$ contributes to a back edge.
    %\item $|P_{in}(\{n\})|> 1$.
\end{enumerate}
A node is \textit{ordinary} if it is not special.
\end{definition}

Figure~\ref{algorithm1} outlines a novel algorithm for abstracting
singly linked lists. The algorithm first collects ordinary nodes of
the input linked list in a set $M$. Then, the algorithm merges any
pair of nodes in $M$ that shares an edge. The merging process
includes adding self-edges. The algorithm supposes that there exists
a function \textit{remove-node} that removes a node from a linked
list.

\begin{figure}[h]
\begin{itemize}
\item[] Algorithm : \textit{Abstract-SLL}
\item[-] Input  :  An abstract component $\hat{C}=(\hat{V},\hat{N},\hat{P})$ such that
$\hat{C}.\lay=\sll$;
\item[-] Output  :  An abstract component $\hat{C}^\prime=(\hat{V}^\prime,
\hat{N}^\prime,\hat{P}^\prime)$ such that $\hat{C}^\prime$ is a
valid abstraction for $\hat{C}$;
\item[-] Method  :\begin{enumerate}
    \item $M\longleftarrow$ ordinary nodes of $\hat{N}$;
    \item While there are $a, b\in M$ such that $a\not= b$ and $(a,b)\in\hat{P}$
   \begin{enumerate}
   \item $(\hat{V},\hat{N},\hat{P})$ = remove-node$(b,\hat{V},\hat{N},\hat{P})$;
   \item $\hat{P}\longleftarrow \hat{P}\cup\{(a,a)\}$;
   \item $M\longleftarrow M\setminus \{b\}$;
   \end{enumerate}
   \item Return $(\hat{V},\hat{N},\hat{P})$;
        \end{enumerate}
\end{itemize}
    \caption{The algorithm \textit{Abstract-SLL}}\label{algorithm1}
\end{figure}

The termination and correctness of \textit{Abstract-SLL} are proved
as follow:

\begin{theorem}\label{theorem1}
The algorithm \textit{Abstract-SLL} always terminates.
\end{theorem}
\begin{proof}
We note that $M$ is finite because $M\subseteq\hat{N}$ and $\hat{N}$
is finite. If the cardinality of $M$ is $m$, then the \textit{while}
loop in the second step iterates at most $m-1$ times.
\end{proof}

\begin{theorem}\label{theorem2}
Suppose that $\hat{C}=(\hat{V},\hat{N},\hat{P})$ is an abstract
component whose layout is $\sll$ and
${\hat{C}^\prime}={\textit{Abstract-SLL}(\hat{C})}$. Then,
$\hat{C}^\prime$ is a valid abstraction of $\hat{C}$.
\end{theorem}
\begin{proof}
We note that there two kinds of operations that occur throughout the
algorithm; removing nodes and adding self-edges. Since both of these
actions do not affect the layout of the component, the layout of the
output component is guaranteed to remain $\sll$. By induction on the
cardinality of $M$, we complete the proof that $\hat{C}^\prime$ is a
valid abstraction of $\hat{C}$. For the induction base, for $|M|=0$
and for $|M|=1$, the required result is trivial. For the inductive
hypothesis, we assume that the required result is true for any
finite set $N$ with $|N| = n$ for some positive integer n. For the
inductive step, we prove the required result holds for a finite set
M with $|M| = n+1$ as follows. We assume that $(a,b)$ is the edge
picked at the first iteration of the loop (if there are none, then
the algorithm terminates and the output is clearly correct). Clearly
$\hat{C}$ is a valid abstraction of itself with the identity witness
$(I_N,I_P)$. Now the component obtained after the first iteration of
the loop, denoted by $\hat{C}^{\prime\prime}$, is a valid
abstraction of $\hat{C}$ with witness
$w^{\prime\prime}=(I_N[a\mapsto b],I_P[(a,b)\mapsto (a,a)])$. The
running of the rest of the algorithm on $\hat{C}$ is equivalent of
that on $\hat{C}^{\prime\prime}$. Clearly $|M|=n$ for the run of
$\hat{C}^{\prime\prime}$. Therefore, by induction hypothesis
$\hat{C}^\prime$ is a valid abstraction of $\hat{C}^{\prime\prime}$
with some witness $w^\prime$. By Lemma~\ref{lem1}, $\hat{C}^\prime$
is a valid abstraction of $\hat{C}$ with witness $w = w^\prime \circ
w^{\prime\prime}$.
\end{proof}

\begin{figure}[htb]
\unitlength.75mm
\begin{center}
\begin{picture}(112,30)(0,0)
\put(-10,0){\oval(7,5)}\put(2,0){\oval(7,5)}\put(14,0){\oval(7,5)}
\put(26,0){\oval(7,5)}\put(38,0){\oval(7,5)}\put(50,0){\oval(7,5)}
\put(62,0){\oval(7,5)}\put(74,0){\oval(7,5)}
\put(-9,1){\makebox[0pt]{\raisebox{-1ex}{{\small $h_7$ }}}}
\put(3,1){\makebox[0pt]{\raisebox{-1ex}{{\small $h_8$ }}}}
\put(15,1){\makebox[0pt]{\raisebox{-1ex}{{\small $h_9$ }}}}
\put(27,1){\makebox[0pt]{\raisebox{-1ex}{{\small $h_{10}$ }}}}
\put(39,1){\makebox[0pt]{\raisebox{-1ex}{{\small $h_{11}$ }}}}
\put(51,1){\makebox[0pt]{\raisebox{-1ex}{{\small $h_{12}$ }}}}
\put(63,1){\makebox[0pt]{\raisebox{-1ex}{{\small $h_{13}$ }}}}
\put(75,1){\makebox[0pt]{\raisebox{-1ex}{{\small $h_{14}$ }}}}
%---------------------------------------------
\put(-3,10){\oval(7,5)}\put(20,10){\oval(7,5)}\put(44,10){\oval(7,5)}
\put(68,10){\oval(7,5)}\put(47.5,10){\vector(1,0){17}}
\put(-3,7.5){\vector(-1,-1){5}}\put(-3,7.5){\vector(1,-1){5}}
\put(20,7.5){\vector(-1,-1){5}}\put(20,7.5){\vector(1,-1){5}}
\put(44,7.5){\vector(-1,-1){5}} \put(44,7.5){\vector(1,-1){5}}
\put(68,7.5){\vector(-1,-1){5}} \put(68,7.5){\vector(1,-1){5}}
\put(-2,11){\makebox[0pt]{\raisebox{-1ex}{{\small $h_3$ }}}}
\put(21,11){\makebox[0pt]{\raisebox{-1ex}{{\small $h_4$ }}}}
\put(45,11){\makebox[0pt]{\raisebox{-1ex}{{\small $h_5$ }}}}
\put(69,11){\makebox[0pt]{\raisebox{-1ex}{{\small $h_6$ }}}}
%---------------------------------------------
\put(8,20){\oval(7,5)}\put(55,20){\oval(7,5)}
\put(8,17.5){\vector(-2,-1){10}} \put(8,17.5){\vector(2,-1){10}}
\put(55,17.5){\vector(-2,-1){10}} \put(55,17.5){\vector(2,-1){10}}
\put(9,21){\makebox[0pt]{\raisebox{-1ex}{{\small $h_1$ }}}}
\put(56,21){\makebox[0pt]{\raisebox{-1ex}{{\small $h_2$ }}}}
%---------------------------------------------
\put(32,30){\oval(7,5)} \put(10,30){\circle{5}}
\put(11,31){\makebox[0pt]{\raisebox{-1ex}{{\small R }}}}
\put(12.5,30){\vector(1,0){16}} \put(32,27.5){\vector(-4,-1){21}}
\put(32,27.5){\vector(4,-1){21}}
\put(33,31){\makebox[0pt]{\raisebox{-1ex}{{\small $h_0$ }}}}
%-------------------------------------------------------------------
%-------------------------------------------------------------------
\put(94,0){\oval(7,5)}\put(106,0){\oval(7,5)}\put(118,0){\oval(7,5)}
\put(130,0){\oval(7,5)}
\put(95,1){\makebox[0pt]{\raisebox{-1ex}{{\small $i_5$ }}}}
\put(107,1){\makebox[0pt]{\raisebox{-1ex}{{\small $i_6$ }}}}
\put(119,1){\makebox[0pt]{\raisebox{-1ex}{{\small $i_7$ }}}}
\put(131,1){\makebox[0pt]{\raisebox{-1ex}{{\small $i_8$ }}}}
%-------------------------------------------------------------------
\put(100,10){\oval(7,5)}\put(124,10){\oval(7,5)}\put(103.5,10){\vector(1,0){17}}
\put(100,7.5){\vector(-1,-1){5}}\put(100,7.5){\vector(1,-1){5}}
\put(124,7.5){\vector(-1,-1){5}}\put(124,7.5){\vector(1,-1){5}}
\put(101,11){\makebox[0pt]{\raisebox{-1ex}{{\small $i_3$ }}}}
\put(125,11){\makebox[0pt]{\raisebox{-1ex}{{\small $i_4$ }}}}
%-------------------------------------------------------------------
\put(90,20){\oval(15,5)}\put(112,20){\oval(7,5)}
\put(84,22.5){{\large $\curvearrowright$}} \put(90,22.5){{\large
$\curvearrowright$}}
\put(87,27){\makebox[0pt]{\raisebox{-1ex}{{\small l }}}}
\put(93,27){\makebox[0pt]{\raisebox{-1ex}{{\small r }}}}
\put(112,17.5){\vector(-2,-1){10}} \put(112,17.5){\vector(2,-1){10}}
\put(91,21){\makebox[0pt]{\raisebox{-1ex}{{\small $i_1$ }}}}
\put(113,21){\makebox[0pt]{\raisebox{-1ex}{{\small $i_2$ }}}}
%-------------------------------------------------------------------
\put(101,30){\oval(7,5)} \put(79,30){\circle{5}}
\put(80,31){\makebox[0pt]{\raisebox{-1ex}{{\small R }}}}
\put(81.5,30){\vector(1,0){16}} \put(101,27.5){\vector(-1,-1){5}}
\put(101,27.5){\vector(2,-1){10}}
\put(102,31){\makebox[0pt]{\raisebox{-1ex}{{\small $i_0$ }}}}
\end{picture}
\end{center}
\caption{A concrete binary tree together with its abstraction.}
\label{bt}
\end{figure}
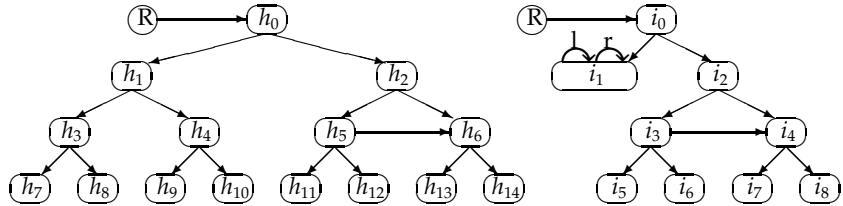

\section{Trees abstraction}\label{bt-s}
In this section, we present a new algorithm for abstracting heap
components whose layout are tree. Figure~\ref{bt} presents a binary
tree (left) of size 15 and height 3 together with its abstracted
representation (right) of size 8 and height 3. We note that node
$h_0$ is special because it is pointed to by the variable $R$. Also
nodes $h_5$ and $h_6$ are special as they share a horizonal edge of
the tree. The special nodes are not grouped in the abstracted
version of the tree. We also note that there is nothing special
about the left subtree. Therefore, the left subtree is grouped in
the node $i_1$ of the compact tree.  The self-edges of $i_1$ model
the fact that $i_1$ represents a full binary subtree.

\begin{definition}
Let $(\hat{V},\hat{N},\hat{P})$ be an abstract component whose
layout is $\tr$. Then, a node $n\in \hat{N}$ is \textit{special} if
\begin{enumerate}
    \item for some $\hat{v}\in \hat{V},\ (\hat{v},n)\in \hat{P}$, or
    \item there exists $(a,b,\_)\in \hat{P}$ such that $a\not=b,\
    a,b \in \hat{N}, \textit{depth}(a)\ge
    \textit{depth}(b)$, and $n\in\{a,b\}$; i.e., $n$ contributes to a
    horizontal or a back edge.
\end{enumerate}
A node is \textit{ordinary} if it is not special.
\end{definition}

Figure~\ref{algorithm2} presents a new algorithm for abstracting
trees. The algorithm first collects ordinary nodes of the input tree
in a set $M$. Then, the algorithm traverses the tree bottom-up,
merging ordinary nodes. The merging process includes adding
self-edges. The algorithm supposes that there is a function
\textit{remove-nodes} that removes a couple of nodes from a tree.

\begin{figure}[h]
\begin{itemize}
\item[] Algorithm : \textit{Abstract-T}
\item[-] Input  :  An abstract component
$\hat{C}=(\hat{V},\hat{N},\hat{P})$ whose layout is \textit{T} and
whose height is denoted by $h$;
\item[-] Output  :   An abstract component $\hat{C}^\prime=(\hat{V}^\prime,
\hat{N}^\prime,\hat{P}^\prime)$ which is a valid abstraction of
$\hat{C}$;
\item[-] Method  :\begin{enumerate}
    \item $M\longleftarrow$ ordinary nodes of $\hat{N}$;
    \item For $(i=h-1;i>0;i--)$
   \begin{enumerate}
   \item While $M$ has distinct elements $a,b,c$
   such that $\textit{depth}(a)=i$ and $(a,b,l),(a,c,r)\in \hat{P}$
   \begin{enumerate}
    \item $(\hat{V},\hat{N},\hat{P})$ = \textit{remove-nodes}$(b,c,\hat{V},\hat{N},\hat{P})$;
    \item $\hat{P}\longleftarrow \hat{P}\cup\{(a,a,l),(a,a,r)\}$;
    \item $M\longleftarrow M\setminus \{b,c\}$;
   \end{enumerate}
\end{enumerate}
\item Return $(\hat{V},\hat{N},\hat{P})$;
\end{enumerate}
\end{itemize}
    \caption{The algorithm \textit{Abstract-T} }\label{algorithm2}
\end{figure}

The proofs of the following two theorems run along similar lines as
those of Theorems~\ref{theorem1} and~\ref{theorem2}, respectively.

\begin{theorem}\label{theorem3}
The algorithm $\textit{Abstract-T}$ always terminates.
\end{theorem}

\begin{theorem}\label{theorem4}
Suppose that $\hat{C}=(\hat{V},\hat{N},\hat{P})$ is an abstract
component whose layout is $\tr$ and
$\hat{C}^\prime=\textit{Abstract-T}(\hat{C})$. Then,
$\hat{C}^\prime$ is a valid abstraction of $\hat{C}$.
\end{theorem}

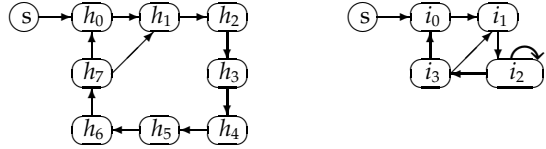
\begin{figure}[htb] \unitlength.75mm
%\center{\input{cycle}}
\begin{center}
\begin{picture}(62,22)(0,0)
\put(-10,20){\circle{5}} \put(-7.5,20){\vector(1,0){6}}
\put(2,10){\oval(7,5)}\put(26,10){\oval(7,5)}
\put(2,0){\oval(7,5)}\put(14,0){\oval(7,5)}\put(26,0){\oval(7,5)}
\put(-9,21){\makebox[0pt]{\raisebox{-1ex}{{\small s }}}}
\put(2,12.5){\vector(0,1){5}} \put(2,2.5){\vector(0,1){5}}
\put(2,20){\oval(7,5)}\put(5.5,20){\vector(1,0){5}}
\put(14,20){\oval(7,5)} \put(17.5,20){\vector(1,0){5}}
\put(26,20){\oval(7,5)}\put(26,17.5){\vector(0,-1){5}}
\put(26,7.5){\vector(0,-1){5}}\put(10.5,0){\vector(-1,0){5}}
\put(22.5,0){\vector(-1,0){5}} \put(5.5,10){\vector(1,1){7.5}}
\put(3,21){\makebox[0pt]{\raisebox{-1ex}{{\small $h_0$ }}}}
\put(15,21){\makebox[0pt]{\raisebox{-1ex}{{\small $h_1$ }}}}
\put(27,21){\makebox[0pt]{\raisebox{-1ex}{{\small $h_2$ }}}}
\put(3,11){\makebox[0pt]{\raisebox{-1ex}{{\small $h_7$ }}}}
\put(27,11){\makebox[0pt]{\raisebox{-1ex}{{\small $h_3$ }}}}
\put(3,1){\makebox[0pt]{\raisebox{-1ex}{{\small $h_6$ }}}}
\put(15,1){\makebox[0pt]{\raisebox{-1ex}{{\small $h_5$ }}}}
\put(27,1){\makebox[0pt]{\raisebox{-1ex}{{\small $h_4$ }}}}
%----------------------------------------------------------------
%----------------------------------------------------------------
\put(50,20){\circle{5}} \put(52.5,20){\vector(1,0){6}}
\put(51,21){\makebox[0pt]{\raisebox{-1ex}{{\small s }}}}
\put(62,20){\oval(7,5)}\put(65.5,20){\vector(1,0){5}}
\put(74,20){\oval(7,5)} \put(74,17.5){\vector(0,-1){5}}
\put(62,10){\oval(7,5)}\put(77,10){\oval(10,5)}
\put(76,12.5){{\large $\curvearrowright$}}
\put(62,12.5){\vector(0,1){5}}\put(65.5,10){\vector(1,1){7.5}}
\put(72,10){\vector(-1,0){6.5}}
\put(63,21){\makebox[0pt]{\raisebox{-1ex}{{\small $i_0$ }}}}
\put(75,21){\makebox[0pt]{\raisebox{-1ex}{{\small $i_1$ }}}}
\put(63,11){\makebox[0pt]{\raisebox{-1ex}{{\small $i_3$ }}}}
\put(78,11){\makebox[0pt]{\raisebox{-1ex}{{\small $i_2$ }}}}
\end{picture}
\end{center}
\caption{A concrete cycle together with its
abstraction.}\label{cycle}
\end{figure}
\section{Cycles abstraction}\label{cycle-s}
The new algorithm presented in this section takes care of
abstracting heap components whose layout is a cycle.
Figure~\ref{cycle} presents a cycle (left) of size 8 together with
its abstracted representation (right) of size 4. We note that node
$h_0$ is special because it is pointed to by the variable $S$. Also
nodes $h_7$ and $h_1$ are special because there are more than one
edge leaving and entering, respectively, the nodes.  As it should
happen, the special nodes are not grouped in the abstracted version
of the cycle. We also note that there is nothing special about nodes
$h_2$ up to node $h_6$. Therefore, these nodes are grouped in the
node $i_2$ of the compressed cycle. The self-edge of $i_2$ models
the fact that $i_2$ represents a sequence of arbitrary length of the
cycle. Special and ordinary nodes of cycles are defined as follows:

\begin{definition}
Let $(\hat{V},\hat{N},\hat{P})$ be an abstract component whose
layout is $\cy$. Then, a node $n\in \hat{N}$ is \textit{special} if
\begin{itemize}
    \item for some $\hat{v}\in \hat{V},\ (\hat{v},n)\in \hat{P}$, or
    \item $|P_{\textit{in}}(\{n\})|>1$, or
    \item $|P_{\textit{out}}(\{n\})|>1$.
\end{itemize}
A node is \textit{ordinary} if it is not special.
\end{definition}

Figure~\ref{algorithm3} presents an original algorithm,
\textit{Abstract-C}, for cycle abstraction. Similar to the
algorithms presented so far, the first step of the algorithm is to
collect ordinary nodes of the cycle. The algorithm then repeatedly
picks a pair of ordinary nodes that share a direct edge. The
algorithm removes one of the two nodes with its edges and adds a
self-node to the remaining node.

\begin{figure}[htb]
\begin{itemize}
\item[] Algorithm: \textit{Abstract-C}
\item[-] Input  :  An abstract component $\hat{C}=(\hat{V},\hat{N},\hat{P})$ such that
$\hat{C}.\lay=\cy$;
\item[-] Output  :  An abstract component $\hat{C}^\prime=(\hat{V}^\prime,
\hat{N}^\prime,\hat{P}^\prime)$ such that $\hat{C}^\prime$ is a
valid abstraction for $\hat{C}$;
\item[-] Method  :\begin{enumerate}
    \item $M\longleftarrow$ ordinary nodes of $\hat{N}$;
    \item While there are $a, b\in M$ such that $a\not= b$ and $(a,b)\in\hat{P}$
   \begin{itemize}
   \item While there exists $(b,c)\in \hat{P}$ \begin{enumerate}
   \item $\hat{P}\longleftarrow (\hat{P}\setminus
   \{(b,c)\})\cup\{(a,c)\}$;
   \end{enumerate}
   \item $\hat{N}\longleftarrow \hat{N}\setminus \{b\}$;
   \item $\hat{P}\longleftarrow (\hat{P}\setminus
   \{(a,b)\})\cup\{(a,a)\}$;
   \item $M\longleftarrow M\setminus \{b\}$;
   \end{itemize}
   \item Return $(\hat{V},\hat{N},\hat{P})$;
        \end{enumerate}
\end{itemize}
    \caption{The algorithm \textit{Abstract-C} }\label{algorithm3}
\end{figure}

The proofs of the following two theorems, which address termination
and correctness of the algorithm \textit{Abstract-C}, are similar to
proofs of Theorems~\ref{theorem1} and~\ref{theorem2}, respectively.

\begin{theorem}
The algorithm \textit{Abstract-C} always terminates.
\end{theorem}

\begin{theorem}
Suppose that $\hat{C}=(\hat{V},\hat{N},\hat{P})$ is an abstract
component whose layout is $\cy$ and
$\hat{C}^\prime=\textit{Abstract-C}(\hat{C})$. Then,
$\hat{C}^\prime$ is a valid abstraction of $\hat{C}$.
\end{theorem}

\begin{figure}[htb]
\unitlength.75mm
\begin{center}
\begin{picture}(112,20)(0,0)
\put(-10,10){\circle{5}}
\put(-9,11){\makebox[0pt]{\raisebox{-1ex}{{\small s }}}}
\put(-7.5,10.5){\vector(1,0){6}} \put(32,20){\oval(7,5)}
\put(29,18){\vector(-4,-1){24}}\put(29,18){\vector(-2,-1){12}}
\put(29,18){\vector(-1,-2){2.7}}\put(33,17.5){\vector(1,-2){2.7}}
\put(33,17.5){\vector(2,-1){13.5}}\put(35,18){\vector(4,-1){24}}
\put(2,10){\oval(7,5)} \put(14,10){\oval(7,5)}
\put(26,10){\oval(7,5)} \put(38,10){\oval(7,5)}
\put(50,10){\oval(7,5)} \put(62,10){\oval(7,5)}
\put(29,2){\vector(-4,1){24}}\put(29,2){\vector(-2,1){12}}
\put(29,2){\vector(-1,2){2.7}}\put(33,2.5){\vector(1,2){2.7}}
\put(33,2.5){\vector(2,1){13.5}}\put(35,2){\vector(4,1){24}}
\put(32,0){\oval(7,5)}
\put(33,21){\makebox[0pt]{\raisebox{-1ex}{{\small $h_0$ }}}}
\put(3,11){\makebox[0pt]{\raisebox{-1ex}{{\small $h_1$ }}}}
\put(15,11){\makebox[0pt]{\raisebox{-1ex}{{\small $h_2$ }}}}
\put(27,11){\makebox[0pt]{\raisebox{-1ex}{{\small $h_3$ }}}}
\put(39,11){\makebox[0pt]{\raisebox{-1ex}{{\small $h_4$ }}}}
\put(51,11){\makebox[0pt]{\raisebox{-1ex}{{\small $h_5$ }}}}
\put(63,11){\makebox[0pt]{\raisebox{-1ex}{{\small $h_6$ }}}}
\put(33,1){\makebox[0pt]{\raisebox{-1ex}{{\small $h_7$ }}}}
%----------------------------------------------------------------
\put(80,10){\circle{5}}
\put(81,11){\makebox[0pt]{\raisebox{-1ex}{{\small s }}}}
\put(82.5,10.5){\vector(1,0){7}}\put(102,20){\oval(7,5)}
\put(99,18){\vector(-1,-2){3}}\put(104,17.5){\vector(1,-1){5}}
\put(93,10){\oval(7,5)}
\put(111,10){\oval(12,5)}\put(110,12.5){{\large $\curvearrowright$}}
\put(99,2){\vector(-1,2){3}}\put(104,2.5){\vector(1,2){2.5}}
\put(102,0){\oval(7,5)}
\put(103,21){\makebox[0pt]{\raisebox{-1ex}{{\small $i_0$ }}}}
\put(94,11){\makebox[0pt]{\raisebox{-1ex}{{\small $i_1$ }}}}
\put(112,11){\makebox[0pt]{\raisebox{-1ex}{{\small $i_2$ }}}}
\put(103,1){\makebox[0pt]{\raisebox{-1ex}{{\small $i_3$ }}}}
\end{picture}
\end{center}
\caption{A concrete DAG together with its abstraction.} \label{DAG}
\end{figure}
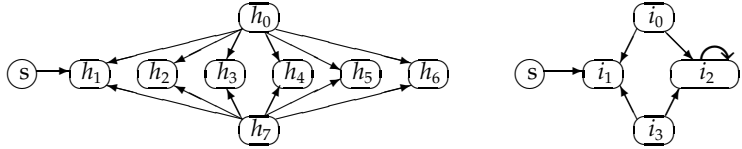
\section{DAG abstraction}\label{dag-s}
This section presents a novel way to abstract heap components whose
layout are \textit{DAG}. Figure~\ref{DAG} presents a \textit{DAG}
(left) of size 8 together with its abstracted representation (right)
of size 4. Node $h_0$ is special because it is pointed to by the
variable $S$. This special node is kept separate in the abstracted
version of the \textit{DAG}. There is nothing special about nodes
$h_2$ up to node $h_6$. Therefore, these nodes are grouped in the
node $i_2$ of the compressed \textit{DAG}. The self-edge of $i_2$
models the fact that $i_2$ represents a set of nodes that is
reference similar. Two distinct nodes of a \textit{DAG} are
reference similar if they are not connected by an edge, point to the
same set of nodes, and are pointed to by the same set of nodes. A
set of nodes is reference similar if every distinct pair of its
elements are reference similar. The following definitions formally
introduce concepts of special nodes, ordinary nodes, and reference
similarity.

\begin{definition}
Let $(\hat{V},\hat{N},\hat{P})$ be an abstract component whose
layout is $\dagg$. Then, a node $n\in \hat{N}$ is \textit{special}
if for some $\hat{v}\in \hat{V},\ (\hat{v},n)\in \hat{P}$. A node is
\textit{ordinary} if it is not special.
\end{definition}

\begin{definition}
Let $\hat{C}=(\hat{V},\hat{N},\hat{P})$ be an abstract component
such that $\hat{C}.\lay=\dagg$. Two distinct nodes $a,b\in\hat{N}$
are \textit{reference similar} with respect to $\hat{C}$ if
\begin{enumerate}
\item $(a,b)\notin \hat{P}$ and $(b,a)\notin \hat{P}$,
\item $\{c\in\hat{N}\mid (c,a)\in\hat{P} \}=\{c\in\hat{N}\mid
(c,b)\in\hat{P}\}$, and
\item $\{c\in\hat{N}\mid (a,c)\in\hat{P} \}=\{c\in\hat{N}\mid (b,c)\in\hat{P}\}$.
\end{enumerate}
A set of nodes $A\subseteq \hat{N}$ is \textit{reference similar}
with respect to $\hat{C}$ if every pair of distinct elements in $A$
is reference similar with respect to $\hat{C}$.
\end{definition}

Figure~\ref{algorithm4} presents the algorithm \textit{Abstract-DAG}
that abstracts heap components with \textit{DAG} layout. The
algorithm calls the algorithm \textit{Ref-similar-DAG},
Figure~\ref{algorithm5}, that for a given heap component calculates
a set of reference similar sets.

\begin{figure}[htb]
\begin{itemize}
\item[] Algorithm : \textit{Abstract-DAG}
\item[-] Input  :  An abstract component $\hat{C}=(\hat{V},\hat{N},\hat{P})$ such that
$\hat{C}.\lay=\dagg$;
\item[-] Output  :  An abstract component $\hat{C}^\prime=(\hat{V}^\prime,
\hat{N}^\prime,\hat{P}^\prime)$ such that $\hat{C}^\prime$ is a
valid abstraction for $\hat{C}$;
\item[-] Method  :\begin{enumerate}
    \item $G^\prime\longleftarrow\textit{Ref-similar-DAG}(\hat{C})$;
    \item $G\longleftarrow\{A\mid A\in G^\prime \mbox{ and }|A|>1
    \}$;
    \item For every $A=\{a_1,a_2,\ldots,a_n\}\in G$
    \begin{enumerate}
    \item $\hat{N}\longleftarrow \hat{N}\setminus \{a_2,\ldots,a_n\}$;
    \item $\hat{P}\longleftarrow \hat{P}\setminus\{(a,b)\mid \{a,b\}\cap\{a_2,\ldots,a_n\}\not=\emptyset\}$;
    \item $\hat{P}\longleftarrow \hat{P}\cup\{(a_1,a_1)\}$;
    \end{enumerate}
\item Return $(\hat{V},\hat{N},\hat{P})$;
\end{enumerate}
\end{itemize}
    \caption{The algorithm \textit{Abstract-DAG} }\label{algorithm4}
\end{figure}

The first step of the algorithm \textit{Abstract-DAG} is to
calculate a set, $G^\prime$, of reference similar sets. The
singleton elements of $G^\prime$ are filtered out to obtain the set
$G$. For every set $A$ in $G$, the algorithm groups elements of $A$
into a single node of the abstracted \textit{DAG} with a self-edge.
The algorithm \textit{Ref-similar-DAG} first initializes $G$ to the
empty set. Secondly, the algorithm stores the ordinary nodes of the
input component in the set $M$. The third step is to partition the
set of ordinary elements, $M$, into reference similar sets. This is
done via picking an element $a\in M$ and adding all elements that
are reference similar to $a$ to the partition of $a$.

\begin{figure}[htb]
\begin{itemize}
\item[] Algorithm : \textit{Ref-similar-DAG}
\item[-] Input  :  An abstract component $\hat{C}=(\hat{V},\hat{N},\hat{P})$ such that
$\hat{C}.\lay=\dagg$;
\item[-] Output  :   A set $G$ of finite subsets of $\hat{N}$ such that
every set of $G$ is reference similar;
\item[-] Method  :\begin{enumerate}
    \item $G\longleftarrow\emptyset$;
    \item $M\longleftarrow$ ordinary elements of $\hat{N}$;
    \item While $M\not= \emptyset$
    \begin{enumerate}
    \item Pick $a$ from $M$.
    \item $A\longleftarrow \{a\}$
    \item For every $b\in M$, \\
          \hspace{1cm} If $A\cup\{b\}$ is reference similar,
          then $A\longleftarrow A\cup\{b\}$;
    \item $G\longleftarrow G\cup\{A\}$;
    \item $M\longleftarrow M\setminus A$;
\end{enumerate}
\item Return $G$;
\end{enumerate}
\end{itemize}
    \caption{The algorithm \textit{Ref-similar-DAG}}\label{algorithm5}
\end{figure}

The proof of the following theorem is similar to that of
Theorem~\ref{theorem1}.

\begin{theorem}
The algorithm \textit{Ref-similar-DAG} always terminates.
\end{theorem}

The proof of the following theorem is by induction on the
cardinality of $M$.

\begin{theorem}
Suppose that $\hat{C}=(\hat{V},\hat{N},\hat{P})$ is an abstract
component and ${G}={\textit{Ref-similar-DAG}(\hat{C})}$. Then, every
element of $G$ is reference similar with respect to $\hat{C}$.
\end{theorem}

\begin{theorem}
The algorithm \textit{Abstract-DAG} always terminates.
\end{theorem}

\begin{theorem}
Suppose that $\hat{C}=(\hat{V},\hat{N},\hat{P})$ is an abstract
component whose layout is \textit{DAG} and
$\hat{C}^\prime=$\textit{Abstract-DAG}$(\hat{C})$. Then,
$\hat{C}^\prime$ is a valid abstraction of $\hat{C}$.
\end{theorem}

\section{Heap abstraction}\label{heap-s}
This section presents our basic algorithm, \textit{Heap-Abstract},
for heap abstraction. For a given abstract heap of $n$ components,
the algorithm checks the layout of each component and calls the
appropriate algorithm for abstracting the component in hand. The
algorithm is outlined in Figure~\ref{algorithm6}. The termination
and correctness of the algorithm are inherited from those of
algorithms presented so far.

\begin{figure}[htb]
\begin{itemize}
\item[] Algorithm : \textit{Heap-Abstract}
\item[-] Input  :  A concrete heap $h=(C_1,\ldots,C_n)$;
\item[-] Output  :  An abstract heap $\hat{h}=(\hat{C}_1,\ldots,\hat{C}_n)$
such that $\hat{h}$ is a valid abstraction for $h$;
\item[-] Method  :\begin{enumerate}
\item For ($i=1;\ i{++};\ i\le n$)
\begin{enumerate}
\item If ($C_i.\lay=\textit{SLL}$), then
$\hat{C}_i\longleftarrow\textit{Abstract-SLL}(\hat{C}_i)$;
\item If ($C_i.\lay=\textit{T}$), then
$\hat{C}_i\longleftarrow\textit{Abstract-T}(\hat{C}_i)$;
\item If ($C_i.\lay=\textit{C}$), then
$\hat{C}_i\longleftarrow\textit{Abstract-C}(\hat{C}_i)$;
\item If ($C_i.\lay=\textit{DAG}$), then
$\hat{C}_i\longleftarrow\textit{Abstract-DAG}(\hat{C}_i)$;
\end{enumerate}
\item Return $(\hat{C}_1,\ldots,\hat{C}_n)$;
\end{enumerate}
\end{itemize}
    \caption{The algorithm \textit{Heap-Abstract} }\label{algorithm6}
\end{figure}

\begin{theorem}
The algorithm \textit{Heap-Abstract} always terminates.
\end{theorem}

\begin{theorem}
Suppose that $h$ is a concrete heap and
$\hat{h}=\textit{Heap-Abstract}(h)$. Then, $\hat{h}$ is a valid
abstraction of $h$.
\end{theorem}

\section{Related work}\label{related work}
The area of statically improving heap allocation, abstraction, and
layout for object oriented programs is rich in
literature~\cite{Cherem06,Dillig10,Magill10,Marron09,Puffitsch10,Vakilian09,Wang10}.
These techniques are conveniently applicable to large programs and
use results of pointer analysis to calculate static partitions that
are required to compute region information. However, there are
common drawbacks to these techniques; (a) they have a limited
capability to conveniently analyze programs that rearrange regions
and (b) they have a limited capability to conveniently explore
components of large complex structures. These deficiencies are
caused by imprecision of determined partitioning and flow
insensitivity. Our algorithms for heap optimization presented in
this paper overcome these drawbacks.

Other techniques that are based on separation logic
~\cite{Berdine07,Guo07} simulates destructive updates of heaps and
how these updates modify heap
layout~\cite{Wilhelm00,Yang08,Beyer10,Jeannet10,El-Zawawy12-2,El-Zawawy11,El-Zawawy11-2,El-Zawawy11-3,El-Zawawy12,El-Zawawy11-4}.
These techniques precisely simulate complicated heap operations but
the limitations imposed by them render these techniques
inappropriate for region analysis. This drawback is witnessed by the
fact that most of these approaches are formulated to analyze
programs that handle only lists or trees. A future direction of
research is to extend the techniques of these papers in the spirit
of our present paper. This is huge potential in this direction by
virtue of generality of separation logic as a general-purpose
framework.

Mathematical domains and maps between domains can be used to
mathematically represent programs and data structures. This
representation is called denotational semantics of programs. One of
our directions for future research is to translate heap concepts to
the side of denotational semantics~\cite{El-Zawawy06,El-Zawawy07}.
Doing so provide a good tool to mathematically study in deep heap
concepts. Then obtained results can be translated back to the side
of programs and data structures.

%\bibliographystyle{unsrt}
%\bibliography{Xbib}
\end{document}